\documentclass{SCIS2024}

\begin{document}
\ArticleType{NEWS \& VIEWS}
\Year{2024}
\Month{}
\Vol{}
\No{}
\DOI{}
\ArtNo{}
\ReceiveDate{}
\ReviseDate{}
\AcceptDate{}
\OnlineDate{}

\title{BUPTCMCC-6G-DataAI+: A generative channel dataset for 6G AI air interface research}{Title for citation}

\author[1]{Li Yu}{}
\author[1]{Jianhua Zhang}{{jhzhang@bupt.edu.cn}}
\author[1]{Mingjun Fu}{}
\author[2]{Qixing Wang}{}

\AuthorMark{Li Yu}

\AuthorCitation{Li Yu, Jianhua Zhang, Mingjun Fu, et al}


\address[1]{State Key Laboratory of Networking and Switching Technology, Beijing University of Posts and Telecommunications, \\Beijing {\rm 100876}, China}
\address[2]{China Mobile Research Institute, Beijing {\rm 100053}, China}

\maketitle

\begin{multicols}{2}
\noindent In September 2024, Beijing University of Posts and Telecommunications and China Mobile Communications Group jointly releases a channel dataset for the sixth generation (6G) mobile communications, named BUPTCMCC-6G-DataAI+. BUPTCMCC-6G-DataAI+ is the update version  of BUPTCMCC-6G-DataAI, which is already published in June 2023, aiming at extending 6G new technologies, frequency bands, and applications.
BUPTCMCC-6G-DataAI+ provides deterministic data covering new mid-bands, millimeter wave (mmWave), and terahertz (THz), supports the features of XL-MIMO near-field, high mobility and provides multiple 6G scenarios such as reconfigurable intelligent surface (RIS) and industrial Internet. Configured with customized features according to different user needs, BUPTCMCC-6G-DataAI+ can adaptively generate scalable large-scale or small-scale parameters, providing data support for 6G research and development, and standardization.

With the global development of 6G research, new technologies, frequency bands, and applications have been proposed~\cite{1}. New frequency bands such as new mid-bands, millimeter wave (mmWave), and terahertz (THz) bring richer spectrum resources and larger bandwidth to 6G communication systems, thereby improving transmission rates. Integrated AI and communication is proposed by ITU-R as a typical application scenario for 6G. 
The connection between AI and communication has become increasingly close, enabling 6G communication algorithms design and optimization, such as beam management for the physical layer, power allocation in the resource layer and network planning and optimization in the network function layer.
With the development of task-oriented AI~\cite{2}, more specified and sufficient training data is required, generative dataset which can produce specific and customized channel parameters with the required data distributions and features is urgently needed.

Some datasets have been constructed for open access, which are generated from statistical models, channel measurements or deterministic scenes. The dataset in Intelligent Wireless Communication Data Public Platform is obtained from channel measurements, which is measured from practical 5G networks in outdoor scenarios\footnote{\href{https://wireless-intelligence.com/\#/download}{https://wireless-intelligence.com/\#/download}}. DeepMIMO~\cite{6} and Wireless AI Research Dataset~\cite{7} are constructed based on ray-tracing data obtained from accurate simulation of deterministic scenes. However, some propagation mechanism, e.g., diffraction is not included in the existing simulation dataset, which is essential for some algorithms design such as long-range blockage prediction~\cite{4}. In addition, 6G new characteristics, like near-field spherical wave feature of EMIMO, RIS, and the sensing scenarios are not considered in the existing datasets. Therefore, BUPTCMCC-6G-DataAI+ is further improved on the basis of BUPTCMCC-DataAI-6G which supports partial features and scenarios, adding new 6G scenarios and features to provide more diversified user-customized 6G channel data for task-oriented AI models.

BUTPCMCC-6G-DataAI+ provides varieties of configuration options, including scenarios, base stations, user points, frequencies, mobile interpolation, enabling RIS and enabling dynamic scenes etc. Based on channel simulation which considers all radio wave propagation mechanisms in constructed scenarios, BUTPCMCC-6G-DataAI+ can process the input parameters and extract corresponding simulation data to generate the required large-scale or small-scale parameters with specific data volume. The massive amounts of customized data provides full coverage in time- frequency- and space domain, supports new features of space-air-ground-sea integration, EMIMO near-field, reconfigurable intelligent surface (RIS), Industrial Internet and high mobility, and covers multiple frequencies including new mid-bands, millimeter wave (mmWave), and terahertz (THz). Since the release of BUTPCMCC-DataAI-6G, it has been downloaded over 2000 times by more than 50 enterprises and universities.

\begin{figure*}[!t]
	\centering
	
	\begin{minipage}{1\linewidth}
		\centering
		\includegraphics[height=7.9cm]{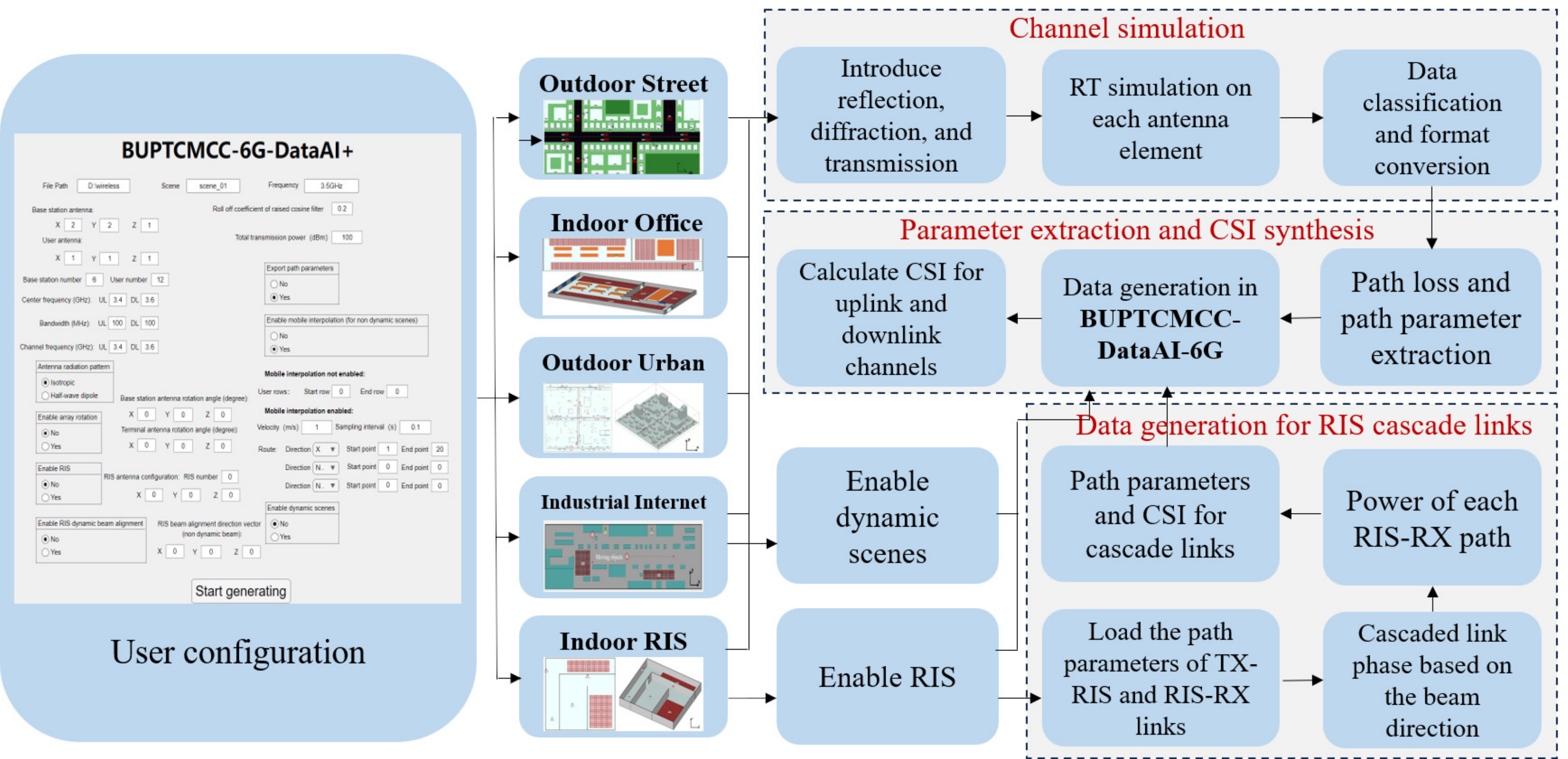}
	\end{minipage}
		
	\caption{ The architecture of BUPTCMCC-6G-DataAI+.
	}
	\label{fig1}
\end{figure*}

\lettersection{Scenarios}
BUPTCMCC-6G-DataAI+ provides multiple scenarios supporting 6G new scenarios. The streets in Outdoor Street Scenario have buildings and vegetation of different heights and sizes on both sides, with a cumulative distribution of 200,000 user points. Indoor Office Scenario is constructed according to the actual office structure, including a large number of common scatterers in the office such as desks, partitions, and conference table. There are obstructions between the transmitter and receiver in Indoor RIS Scenario, where RIS boards are equipped, providing TX-RIS-RX cascade channels. The Industrial Internet Scenario is constructed according to the actual factory scenarios, including the typical features of the Industrial Internet. The Outdoor Urban Scenario is rich in environment and contains many kinds of houses and vehicles.

\lettersection{Functions}
BUTPCMCC-6G-DataAI+ allows input parameter flexible configuration and specified channel parameter output. In the dataset interface, users can flexibly specify parameters such as base station, user areas, antennas, scenarios and frequencies and choose whether to enable mobile interpolation, RIS and dynamic scenes. In addition to the output of CSI, output of path parameters is also allowed, further enriching the generated data. Using the point interpolation method, the user's moving velocity and direction can be freely configured~\cite{5}, breaking the limitation of fixed simulation points in the traditional simulation channel dataset. In the Indoor RIS Scenario, the RIS boards supports the beam adjustment, which allows the user to customize the direction of beam alignment. In the Industrial Internet Scenario, dynamic scene data is provided. In order to simulate the motion of automation equipments such as automatic transportation trolleys, two dynamic objects are set up to move along a specified route.

\lettersection{Applications}
BUTPCMCC-6G-DataAI+ provides diversified data support for 6G wireless communication system, where AI enables end-to-end output and transmission solutions for different layers.
Based on the generated large-scale parameters, overall system planning can implemented, determining the location deployment of base stations and the allocation of spectrum resources while small-scale parameters contain information that determines the number of signal streams transmitted from the base station to the user, the precoding matrix of each signal stream, and power allocation.

Traditional datasets provide fixed combinations of data and cannot be configured according to user needs to provide personalized data, while diversified user-customized 6G channel data can be generated from BUTPCMCC-6G-DataAI+ to enable task-oriented AI models for deeper integration of AI and communication.

BUPTCMCC-6G-DataAI+ is available on the High Performance Computing Platform of Beijing University of Posts and Telecommunications\footnote{\href{https://hpc.bupt.edu.cn/dataset-manage/home-page/}{https://hpc.bupt.edu.cn/dataset-manage/home-page}} and China Mobile Smart Network New Generation Artificial Intelligence Open Innovation Platform\footnote{\href{https://jiutian.10086.cn/open/\#/firstpage?platform=OpenInnovation}{https://jiutian.10086.cn/open/\#/firstpage?platform=OpenInnovation}}, where detailed information can be found.

\lettersection{Views}
In the future, further development and evolution will emerge at the 6G technology, inducing more intricate appearing of scenarios, features and techniques, where our dataset will be constantly optimized to comprehensively support the profound integration of AI and communication, technological innovation and development of 6G, as well as its future design and eventual standardization and implementation in 6G.




\end{multicols}
\end{document}